\documentclass[11pt]{article}
\usepackage[sc]{mathpazo} 
\usepackage{fullpage}
\usepackage{cite}
\usepackage[utf8]{inputenc}
\usepackage{lineno}
\usepackage{titlesec}
\usepackage[]{graphicx}
\usepackage{float}
\usepackage{color,ulem}
\usepackage{amsmath}

\titleformat{\section}[block]{\Large\bfseries\filcenter}{\thesection}{1em}{}
\titleformat{\subsection}[block]{\Large\itshape\filcenter}{\thesubsection}{1em}{}
\titleformat{\subsubsection}[block]{\large\itshape}{\thesubsubsection}{1em}{}
\titleformat{\paragraph}[runin]{\itshape}{\theparagraph}{1em}{}[. ]

\title{Dynamic trait distribution as a source for shifts in interaction strength and population density}
\author{Zachary Jackson$^{1,\ast}$ \\ 
BingKan Xue$^{1,2}$}
\date{1. Department of Physics, University of Florida, Gainesville, FL, 32611 \\
2. Institute for Fundamental Theory, University of Florida, Gainesville, FL 32611 \\
$\ast$ Corresponding author; e-mail: zackljackson@ufl.edu.}

\date{\footnotesize\textsuperscript{\textbf{1}} Department of Physics, University of Florida, Gainesville, FL, United States\\ \textsuperscript{\textbf{2}}Institute for Fundamental Theory, University of Florida, Gainesville, FL, United States}

\begin{document}

\maketitle

\begin{abstract}

Intraspecific trait variation has been increasingly recognized as an important factor in determining species interaction and diversity. Eco-evolutionary models have studied the distribution of trait values within a population that changes over the generations as a result of selection and heritability. Non-heritable traits that can change within the lifetime, such as behavior, can cause trait-mediated indirect effects, often studied by modeling the dynamics of a homogeneous trait. Complementary to these two approaches, we study the distribution of traits within a population and its dynamics on short timescales due to ecological processes. We consider several mechanisms by which the trait distribution can shift dynamically: phenotypic plasticity within each individual, differential growth among individuals, and preferential consumption by the predator. Through a simple predator-prey model that explicitly tracks the trait distribution within the prey, we illustrate some unusual consequences of trait shifts, including the healthy herd effect and the emergent promotion of the prey by the predator. We analyze the cause of these phenomena by identifying the density and trait effects from the predator. Our results show that intraspecific trait variation and the dynamics of its distribution can lead to the modification of species interactions and result in outcomes that are otherwise unexpected.

\end{abstract}

\maketitle

\section*{Introduction}

In ecological modeling, it is common to use species as the basic unit of consideration, such as using the overall density of a species as a dynamical variable in Lotka-Volttera equations. However, trait differences within a species cannot always be ignored. It has been increasingly recognized that there is significant variation in ecologically important traits among individuals of the same population \cite{westerband:2021, moran:2016}. Such variation can affect how individuals interact with other organisms and their environment, owing to their unique trait makeup \cite{holt:2012, bolnick:2011}. It has been shown that intraspecific trait variation plays an important role in community ecology \cite{bolnick:2011, barabas:2016, hart:2016}. The nonlinear averaging over intraspecific variation may affect the overall interaction between species \cite{okuyama:2008, hart:2016}; in turn, species interactions can influence the trait variation within a species.

A successful framework has been developed to describe quantitative traits and the changes in their distributions in the case of heritable traits \cite{lush:1937, price:1970, lande:1976, heywood:2005}. In these models, the distribution of trait values slowly shifts towards higher fitness through selection over many generations. This framework has recently been combined with shorter timescale population dynamics through an eco-evolutionary feedback \cite{ferriere:2015, govaert:2019}. Such work has mostly considered heritable traits that are normally distributed \cite{pastore:2021,schreiber:2011}. On the other hand, traits that are not necessarily heritable can vary on a timescale shorter than a generation, such as plastic traits in response to environmental variability or developmental stochasticity. In particular, behavioral traits can change quickly in response to the presence of predators and can affect how the prey consume resources \cite{peacor:2001}. Such trait changes and their consequences on ecological dynamics have been studied under the subject of trait-mediated indirect effects \cite{holt:2012, peers:2018, terry:2017}. Previous work has modeled the adaptive dynamics of a homogeneous trait \cite{barfield:2012, abrams:2009}. Meanwhile, interesting and sometimes unintuitive results have been found in ecosystems arising from population structures within species \cite{deroos:2008}. Non-normal trait distributions, such as bimodal distributions, can have strong effects on the dynamics and equilibrium of populations \cite{coutinho:2016}. Our goal here is to incorporate such ``trait structures'' of a population (i.e., the structure of the trait distribution within a population) into the modeling of ecological dynamics to study the consequences of short-timescale shift of trait distributions.

\begin{figure}
\centering
\includegraphics[width=.97\textwidth]{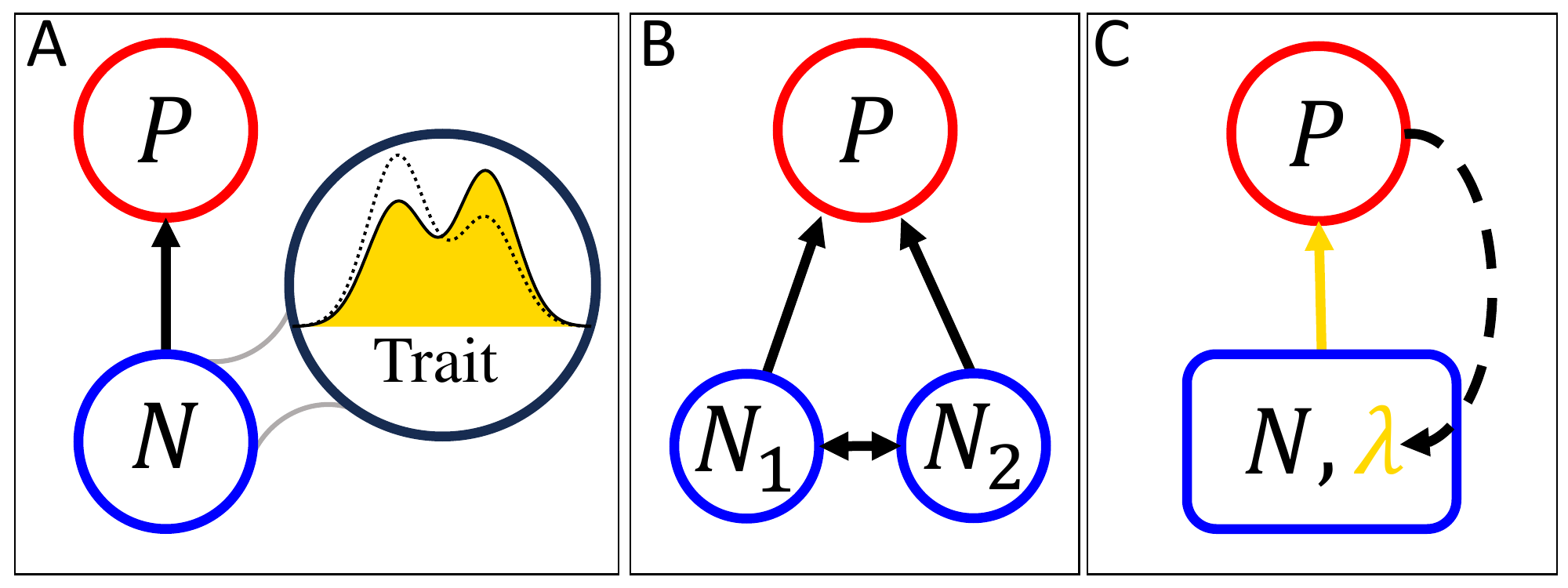}
\caption{(A) The interaction between a predator ($P$) and prey ($N$) can be influenced by the trait variation within the prey population. The distribution of trait values (inset) can shift dynamically due to multiple processes. (B) A simplified model of trait distribution with two categories of prey ($N_1$ and $N_2$), consumed by the predator at different rates. There is a flow between the categories through changes in trait values within or across generations, in response to the predator, environment, or developmental stochasticity. (C) A coarse-grained picture where $\lambda$ represents the trait distribution of the prey. The overall interaction strength between the species depends on the trait distribution (solid arrow), and the trait distribution is in turn affected by the interaction (dashed arrow).}
\label{fig:sketch}
\end{figure}

Consider a species whose individuals form a distribution over a trait space, as represented in Figure~\ref{fig:sketch}A (inset). These traits could be the developmental stage, body size, morphological patterns, behavioral modes, or other properties that can vary from individual to individual and can affect how individuals interact with others. The trait distribution can shift on ecological timescales due to multiple kinds of influences. The trait value may be correlated with the growth rate or reproductive success of individuals. The trait distribution may thus be selected to skew towards trait values associated with faster growth. Besides, predators may disproportionately consume certain types of individuals because they are more enticing or easier to catch, such as due to differences in chemical defense or behavior. The trait distribution of the prey will be affected by this preference because the consumption by the predator will shift the distribution away from what is consumed more of. Moreover, the trait value of an individual can change over its lifetime, such as by phenotypic plasticity in response to environmental signals or by switching into a different behavioral mode, thus shifting the trait distribution. These types of ``trait shifts'' would induce a modification of the overall growth rate and interaction strength of the species. We will study the consequences of such interaction modification due to trait shifts.

To develop a formalism for analyzing the effect of dynamic shift in trait distributions, we consider a minimal ecosystem of one predator and one prey. We will explicitly track the trait distribution within the prey population (Figure~\ref{fig:sketch}A). The trait value of the prey is linked to both the reproduction rate of the prey and the consumption rate by the predator. By allowing the trait distribution to shift dynamically, we show that there is a significant correction to the equilibrium of the system compared to what would otherwise be expected without the trait shift. In particular, we observe the \textit{emergent promotion} of the prey by the predator, whereby the prey density increases in the presence of the predator as compared to in its absence. This phenomenon was found previously in a model of two predators and one prey \cite{jackson:2023}. Here it happens in an even simpler system with only two species, which allows us to trace its cause by dividing the effect of the predator into density and trait effects. We show that unintuitive results arise when the trait effect dominates over the density effect. In the situation of an infectious disease, for example, we see the emergent promotion of prey population by its predator in addition to the ``healthy herd effect'' \cite{packer:2003}. Such phenomena can be considered a trait-mediated \textit{direct} effect, as it influences the trait-shifted species itself, without involving a third species or resource at a lower trophic level.

\section*{Methods}

We start with Lotka-Volterra equations for a predator-prey system including the predator $P$ and the prey $N$. For simplicity, we only consider trait variation within the prey. Furthermore, we consider a discrete trait that can take one of two values, such that the trait distribution is given by a categorical distribution with two categories. This could be an approximate description of trait distributions that are bimodal and the relative abundance of the two modes can change over time. These modes may represent behavioral modes where individual animals are either foraging for food or hiding from predators \cite{preisser:2005, herberholz:2012, bracis:2021}, or physical conditions where individuals are either healthy or infected with a disease \cite{belden:2007,lafferty:2017}. In such cases, the prey population can be divided according to their trait values into two subpopulations, $N_1$ and $N_2$ (Figure~\ref{fig:sketch}B). We allow these subpopulations to have different birth rates ($\nu_1$ and $\nu_2$), death rates ($\delta_1$ and $\delta_2$), and rates of consumption by the predator ($\chi_1$ and $\chi_2$). In addition, there is a flow of individuals between the subpopulations, which represents either individuals changing trait values during their lifetime or individuals being born with a different trait from their parents. We will consider several different scenarios below, and will show that these scenarios can be described in a unified mathematical form.

\subsection*{Intimidated prey}

The first scenario is one where the presence of the predator has a significant influence on the behavior of the prey. This situation has been studied as a trait-mediated indirect effect, such as in a system of grasshoppers and predatory spiders where the grasshoppers modified their feeding-time budget and dietary choices to avoid the spiders \cite{schmitz:1997, beckerman:1997}. We will consider different behavioral modes, such as hiding and foraging, as our trait values that distinguish between the subpopulations of prey. The predator may consume the prey with certain behavior more readily, and the prey may react to the predator by switching its behavior. The dynamical system of the predator and prey is represented by the following equations:
\begin{align*}
\dot{P} &= -\mu P + \varepsilon \chi_1 P N_1 + \varepsilon \chi_2 P N_2 \\
\dot{N_1} &= \nu_1 N_1 + \nu_2 N_2 - \delta_1 N_1 - \alpha N N_1 - \chi_1 P N_1 - \theta(P) N_1 + \phi N_2 \\
\dot{N_2} &= - \delta_2 N_2 - \alpha N N_2 - \chi_2 P N_2 + \theta(P) N_1 - \phi N_2
\end{align*}
The equation for the predator ($P$) includes a death rate $\mu$ and a consumption term for each prey type; $\varepsilon$ represents the efficiency by which consumed prey is converted into predator biomass. There are two equations for the two prey types; we think of $N_1$ as the ``sensitive'' type (which has not yet been intimidated by the predator) and $N_2$ as the ``intimidated'' type. The prey equations have terms for birth, death, intraspecific competition (with the same strength $\alpha$ for both types), and consumption by the predator. In addition, there are switching terms between the subpopulations. We assume that all individuals are born sensitive and only switch to being intimidated when they encounter the predator. The switching rate is represented by $\theta(P)$, which depends on the predator density. Specifically, we assume $\theta(P)$ is such that sensitive individuals switch to being intimidated at a positive rate if such switching increases the overall prey growth. The functional form of $\theta(P)$ is shown in Table~\ref{Table:parameters}. We also assume that the intimidated individuals will relax to being sensitive at a constant rate $\phi$.

\subsection*{Inherent heterogeneity}

The second scenario we consider is one where the two prey types are born in a constant ratio regardless of the parental trait. We referred to this model as ``inherent heterogeneity'' in \cite{jackson:2023}, which assumes that individuals cannot change traits in later stages of life. For example, certain morphological traits are determined during development and do not change later, such as small or large horns in dimorphic beetles \cite{emlen:2005}. The dynamical system to model this scenario is given by the equations:
\begin{align*}
\dot{P} &= -\mu P + \varepsilon \chi_1 P N_1 + \varepsilon \chi_2 P N_2 \\
\dot{N_1} &= (1-\rho) (\nu_1 N_1 + \nu_2 N_2) - \delta_1 N_1 - \alpha N N_1 - \chi_1 P N_1\\
\dot{N_2} &= \rho (\nu_1 N_1 + \nu_2 N_2) - \delta_2 N_2 - \alpha N N_2 - \chi_2 P N_2
\end{align*}
Here $\rho$ and $(1-\rho)$ are the fractions of prey that are born as type 2 and type 1, respectively. We consider type 1 as the ``superior'' type, and type 2 as the ``inferior'' type, because we choose type 2 to be the trait associated with a slower growth rate, without loss of generality. The other terms in the prey equations, as well as the equation for the predator, are the same as for the intimidated prey model above. The difference is that, in the current scenario, the prey does not actively respond to the predator, so there are no switching terms. However, we will see that these equations can be transformed mathematically to have effective switching terms that push the trait distribution of the prey towards the natural composition $\rho$.

\subsection*{Infectious disease}

Lastly, we consider a scenario where an infectious disease spreads among the prey, which separates it into a ``susceptible'' subpopulation ($N_1$) and an ``infected'' subpopulation ($N_2$). Following the Susceptible-Infected (SI) model similar to \cite{hall:2005, packer:2003}, we assume that the infected individuals recover at a constant rate, but are susceptible to reinfection. We also assume that all individuals are born susceptible, i.e., the disease is not vertically transmitted. In addition, the consumption rate by the predator depends on the prey's infection status \cite{hudson:1992}. The dynamical system for this scenario is given by the equations:
\begin{align*}
\dot{P} &= -\mu P + \varepsilon \chi_1 P N_1 + \varepsilon \chi_2 P N_2 \\
\dot{N_1} &= \nu_1 N_1 + \nu_2 N_2 - \delta_1 N_1 - \alpha N N_1 - \chi_1 P N_1 - \beta N_1 N_2 + \gamma N_2 \\
\dot{N_2} &= - \delta_2 N_2 - \alpha N N_2 - \chi_2 P N_2 + \beta N_1 N_2 - \gamma N_2
\end{align*}
Here, infection is represented by the density-dependent term $\beta N_1 N_2$ with $\beta$ being the infection constant, and recovery is represented by the term $\gamma N_2$ with $\gamma$ being the recovery rate. The other terms in the equations are the same as for the intimidated prey model above. Note that in this infection model, the switching of prey types (i.e., the infection and recovery terms) depends on the prey population, instead of depending on the predator as in the intimidated prey model.

\subsection*{Reformulation of the models}

Although the three models above describe very different mechanisms for intraspecific trait variation, they can be put in a common form by redefining some of the parameters. The general set of equations can be written as:
\begin{align*}
\dot{P} &= (-\mu + \varepsilon \chi_1 N_1 + \varepsilon \chi_2 N_2) P \\
\dot{N_1} &= (\nu_1 - \delta_1) N_1 + (\nu_2 - \delta_2) N_2 - \alpha N N_1 - \chi_1 P N_1 - \sigma_1(P,N_i) N_1 + \sigma_2(P,N_i) N_2 \\
\dot{N_2} &= - \alpha N N_2 - \chi_2 P N_2 + \sigma_1(P,N_i) N_1 - \sigma_2(P,N_i) N_2
\end{align*}
The meanings of all the parameters are given in Table~\ref{Table:parameters}. The $\sigma_i$ terms effectively represent switching of the prey from one type to the other, as they appear in both equations of $N_i$ with opposite signs. These terms do not affect the total population of the prey, but can change its trait distribution. The specific forms of these terms for each model are listed in Table~\ref{Table:parameters}. Note that even in the model of inherent heterogeneity, where individuals do not change their phenotype, there are still effective switching terms coming from individuals dying and being replaced with the other type.

\begin{table}[ht]
\caption{Parameters in the models}
\label{Table:parameters}
\centering
\begin{tabular}{ll}\hline
Parameter        &    Meaning                  \\ \hline \hline
$\mu$            &   predator death rate       \\
$\varepsilon$    &   predator conversion efficiency     \\
$\chi_i$       &   predator consumption rate          \\
$\nu_i$          &   prey reproduction rate           \\
$\delta_i$       &   prey death rate           \\
$\alpha$       &   prey intraspecific competition \\ \hline
Intimidated prey    & $\sigma_1 = \text{Max}[0, \eta (P (\chi_1 - \chi_2) - (\nu_1 - \delta_1) + (\nu_2 - \delta_2))]$, \, $\sigma_2 = \delta_2 + \phi$ \\ \hline
$\eta$           &   prey reaction strength    \\
$\phi$         &   prey relaxation rate            \\ \hline  
Inherent heterogeneity  & $\sigma_1 = \rho \nu_1$, \, $\sigma_2 = \delta_2 - \rho \nu_2$   \\ \hline
$\rho$           &   prey birth ratio               \\ \hline
Infectious disease  & $\sigma_1 = \beta N_2$, \, $\sigma_2 = \delta_2 + \gamma$     \\ \hline
$\beta$          &   prey infection constant            \\
$\gamma$         &   prey recovery rate             \\ \hline
\end{tabular}
\end{table}

To focus on the trait distribution of the prey, we further transform these equations in terms of the total prey population $N = N_1 + N_2$ and the fraction $\lambda = N_2 / N$ (Figure~\ref{fig:sketch}C). We also nondimensionalize the equations with the following transformation of variables and definition of parameters:
\begin{align*}
& P \leftarrow \frac{\alpha}{\varepsilon (\nu_1-\delta_1)} P, \quad N \leftarrow \frac{\alpha}{\nu_1-\delta_1}N, \quad t \leftarrow (\nu_1-\delta_1) t \\
& m \equiv \frac{\mu}{\nu_1-\delta_1}, \quad c_1 \equiv \frac{\varepsilon \chi_{1}}{\alpha}, \quad c_2 \equiv \frac{\varepsilon \chi_{2}}{\alpha}, \quad r \equiv 1-\frac{\nu_2-\delta_2}{\nu_1-\delta_1}
\end{align*}
The resulting equations for the predator-prey system are
\begin{align}
\dot{P} &= \big( - m + c_\textrm{eff}(\lambda) N \big) P \label{eq:pdot}\\
\dot{N} &= \big( b_\textrm{eff}(\lambda) - N - c_\textrm{eff}(\lambda) P \big) N \label{eq:ndot}\\
\dot{\lambda} &= \lambda (1-\lambda) P (c_1 - c_2) - \lambda b_\textrm{eff}(\lambda) + (1-\lambda) S_1(P,N,\lambda) - \lambda S_2(P,N,\lambda) \label{eq:lambda-dot}
\end{align}
where $c_\textrm{eff}(\lambda) = (1-\lambda) c_1 + \lambda c_2$ and $b_\textrm{eff}(\lambda) = 1 - r \lambda$. The functional forms of the $S_i$'s are shown in Table 2.

\begin{table}
\caption{Transformed switching terms for different models}
\label{Table:switching}
\centering
\begin{tabular}{lll}\hline
Model        &  $S_1$       & $S_2$            \\ \hline
Intimidated prey  & $\text{Max}[0,u (P (c_1 - c_2) - r)]$ & $d+f$ \\
Inherent heterogeneity      & $\rho b$      & $(1-\rho) d - \rho (1-r)$            \\
Infectious disease     & $q (1-\lambda) N$  & $d + g$          \\ \hline
\end{tabular} \\[6pt]
\raggedright
where $b=\nu_1/(\nu_1-\delta_1)$, $d=\delta_2/(\nu_1-\delta_1)$, $u=\eta/(\nu_1-\delta_1)$, $q=\beta/(\nu_1-\delta_1)$, $f=\phi/(\nu_1-\delta_1)$, $g=\gamma/(\nu_1-\delta_1)$.
\end{table}

If the trait distribution cannot shift, i.e., $\lambda$ is a constant (ignoring Eq.~(\ref{eq:lambda-dot})), then Eqs.~(\ref{eq:pdot}--\ref{eq:ndot}) reduce to the classic Lotka-Volterra equations, with $c_\textrm{eff}$ being the consumption rate of the predator (biomass conversion efficiency normalized to 1), $b_\textrm{eff}$ being the growth rate of the prey, and with a carrying capacity for the prey (normalized to 1). Therefore, the effect of trait shift is captured by the dynamics of $\lambda$ in Eq.~(\ref{eq:lambda-dot}) and how it changes the effective interaction strength $c_\textrm{eff}$ and growth rate $b_\textrm{eff}$. Among the four terms in Eq.~(\ref{eq:lambda-dot}), the first term is the shift in the prey distribution due to consumption by the predator. The trait distribution shifts away from the preference of the predator because the predator consumes more of what it prefers. The second term comes from the difference in growth rate associated with different trait values. The distribution shifts towards the trait value that allows individuals to grow faster. The third and fourth terms are the switching terms that depend on the specific model. Thus, the shape of the trait distribution can be modified by a combination of preferential consumption, differential growth, and effective switching.

\subsection*{Comparison of density and trait effects}

To find the effect of trait shift, we will compare the behavior of the full system to a model with a fixed trait distribution. We can separate the ``trait effect'' from the ``density effect'' that both arise from introducing the predator to a system with only the prey. When the system consists of only the prey, the equilibrium prey population is $N_0$ and the natural trait distribution is $\lambda_0$. When the predator is introduced, the new equilibrium prey population is $N^*$ and the new trait distribution is $\lambda^*$. We will denote by $N^*(\lambda^*)$ the equilibrium prey population in the presence of the predator (when the trait distribution is $\lambda^*$), and $N_0(\lambda_0)$ the equilibrium prey population without the predator (when the trait distribution is $\lambda_0$). Introducing the predator into the system causes a trait shift in the prey, $\Delta\lambda = \lambda^* - \lambda_0$, as well as a change in the density of the prey, $\Delta N = N^* - N_0$. We will divide this prey density change into a density effect and a trait effect from the predator, $\Delta N = \Delta N_D + \Delta N_T$, as follows.

In order to isolate the density effect, we introduce $N^*(\lambda_0)$ and $P^*(\lambda_0)$, which are the hypothetical populations at equilibrium when the trait distribution of the prey is held at its natural value $\lambda_0$. Mathematically, these are given by solving Eqs.~(\ref{eq:pdot}--\ref{eq:ndot}) with a fixed $\lambda$:
\begin{equation}
N^*(\lambda) = \frac{m}{c_\textrm{eff}(\lambda)} \,, \quad
P^*(\lambda) = \frac{b_\textrm{eff}(\lambda) - m / c_\textrm{eff}(\lambda)}{c_\textrm{eff}(\lambda)}
\end{equation}
In the absence of the trait shift, the density effect can be defined as the change in prey population due to the introduction of the predator, i.e., $\Delta N_D = N^*(\lambda_0) - N_0(\lambda_0)$. Then the trait effect is the change in prey population due to the trait shift, i.e., $\Delta N_T = N^*(\lambda^*) - N^*(\lambda_0)$. Similarly, the trait shift of the prey has a ``back-reaction'' on the predator, i.e., $\Delta P_T = P^*(\lambda^*) - P^*(\lambda_0)$. This back-reaction measures how the trait shift caused by the predator affects the population of the predator itself. It is important to note that, because $\Delta P_T$ is a back-reaction from the prey, changes in both prey population and trait distribution are involved. We will examine these different effects to understand the behavior of the predator-prey system.

\section*{Results}

\begin{figure}
\centering
\includegraphics[width=.99\textwidth]{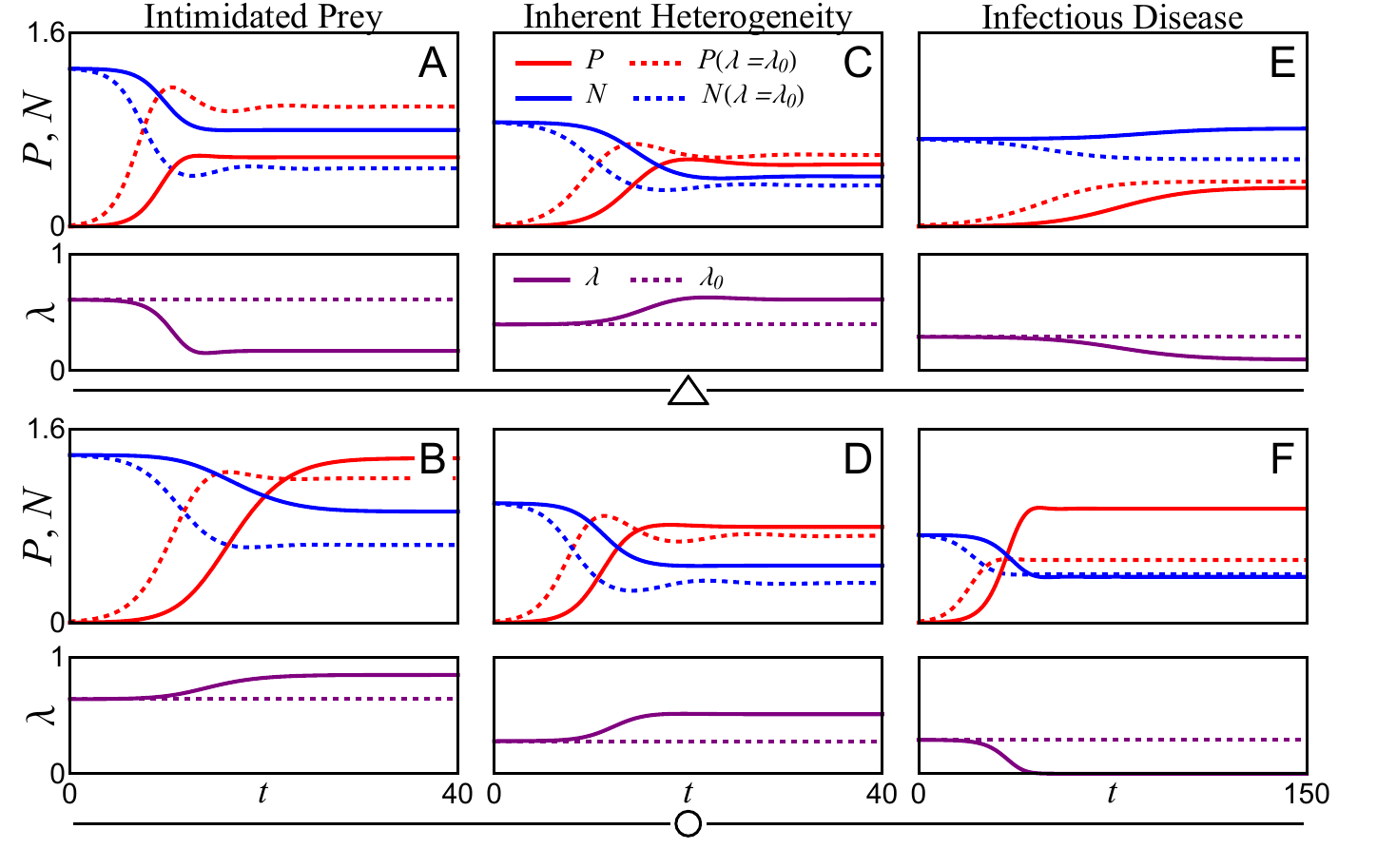}
\caption{Example trajectories of the three models showing different behaviors depending on the prey growth rate difference and predator consumption preference. Panels (A,C,E) correspond to triangles in Figures~\ref{fig:phase_int}-\ref{fig:phase_inf}, and panels (B,D,F) correspond to circles in Figures~\ref{fig:phase_int}-\ref{fig:phase_inf}. Parameter values used for the intimidated prey model are: $(c, d, m, u, f) = (1.5, 0.1, 0.4, 5, 0.2)$; $(x, r) = (0.25, -0.5)$ in panel (A), and $(x, r) = (0.8, -0.6)$ in (B). For the inherent heterogeneity model: $(c, d, m, \rho, b) = (1.3, 0.1, 0.3, 0.3, 1.0)$; $(x, r) = (0.9, 0.5)$ in (C), and $(x, r) = (0.97, 0.05)$ in (D). For the infectious disease model: $(c, d, m, q, g) = (1.2, 0.1, 0.25, 1.8, 0.1)$; $(x, r) = (0.2, 0.95)$ in (E), and $(x, r) = (0.55, 0.95)$ in (F).}
\label{fig:time_all}
\end{figure}

Our reformulation of the three different models allows us to analyze their behavior using the same method. Example trajectories of the dynamical systems (Eqs.~\ref{eq:pdot}--\ref{eq:lambda-dot}) are shown in Figure~\ref{fig:time_all}. Each time series starts with the system at equilibrium with only the prey present. The predator is introduced and the system is allowed to equilibrate (solid curves). We also consider the hypothetical situation where the prey trait distribution is held at its natural value (dashed curves).

Figure~\ref{fig:time_all}AB shows a pair of time series for the intimidated prey model. The prey reacts to the predator by switching behavioral modes depending on the predator's feeding preference. In the first time series (Figure~\ref{fig:time_all}A), the predator prefers to consume the intimidated type. As a result, the composition of the prey shifts away from its preference, which can be seen from the solid purple curve that shows a decreased $\lambda$. This trait shift in the prey has a negative back-reaction on the predator population, as shown by the fact that the equilibrium value reached by the solid red curve is lower than that reached by the dashed red curve. In the second time series (Figure~\ref{fig:time_all}B) where the predator prefers the sensitive type, however, the trait shift helps more prey survive and in turn increases the predator population. This can be seen by noticing that the equilibrium reached by the solid lines are higher than that reached by the dashed lines of each color. Thus, even though in both cases the trait shift goes against the predator's preference, the back-reaction on the predator can be either negative or positive.

Figure~\ref{fig:time_all}CD shows a similar story for the inherent heterogeneity model, where the prey are born into either type and do not react to the predator. In this model the trait shift is caused solely by the predator's preferential consumption of the prey types. Two examples are shown, where the growth rate difference between the two prey types is more pronounced in the first example. In both cases the shift in prey composition is away from the predator's preference. We see once again that the predator population can be decreased (Figure~\ref{fig:time_all}C) or increased (Figure~\ref{fig:time_all}D) by the shift in prey composition: The solid red line representing the predator population reaches an equilibrium that is below the dashed red line in Figure~\ref{fig:time_all}C but above the dashed red line in Figure~\ref{fig:time_all}D.

Figure~\ref{fig:time_all}EF shows example time series for the infectious disease model. Here, the trait distribution $\lambda$ is the proportion of the population infected with the disease. In both of the examples shown in the figure, introducing the predator relieves the prey of the disease, as $\lambda$ goes down. This is true even for the case where the predator slightly prefers susceptible prey (Figure~\ref{fig:time_all}F), because the prey population is reduced such that the spread of the disease is suppressed. On the other hand, the effect on the prey population is very different between the two examples. In Figure~\ref{fig:time_all}F, introducing the predator causes the prey population to fall drastically, but the shift in the prey composition has almost no effect on the equilibrium prey population (solid and dashed blue curves). However, in Figure~\ref{fig:time_all}E, when the predator is introduced, the prey population decreases if the trait composition is held fixed (dashed blue curve), but increases if the composition is allowed to shift (solid blue curve). This is an example of emergent promotion, which we explore in more detail below.

These examples already illustrate some nontrivial outcomes of dynamic trait shifts, including the emergent promotion of the prey and the back reaction on the predator. To explore the generality of those examples, we will carry out a systematic study of the parameter space below. The different dynamics among the similar systems can be attributed to different combinations of density and trait effects.

\subsection*{Phase diagrams}

We examine the systems over the parameter space of the growth difference between the prey types and the feeding preference of the predator. The difference in growth rate between the two prey types is captured by the parameter $r$. When $r = 0$, the growth rates of both prey types are equal; when $r < 0$, the growth rate of type 2 prey is larger than type 1, and vice versa. In the intimidated prey model, we explore $r$ in the range from $-1$ to $1$; in the inherent heterogeneity model, we explore $r$ from $0$ to $1$, because we have assumed type 1 to grow faster without loss of generality; and in the infectious disease model, we also explore $r$ from $0$ to $1$, because a disease tends to reduce the growth rate of the prey. To capture the preference of the predator, we set $c_1 = x \, c$ and $c_2 = (1-x) c$, so that $0 \leq x \leq 1$ controls how strongly the predator prefers one type of prey over the other. When $x \to 0$, the predator strongly prefers type 2 prey; when $x \to 1$, the predator strongly prefers type 1 prey; and when $x = 0.5$, the predator has no preference.

\begin{figure}
\centering
\includegraphics[width=.95\textwidth]{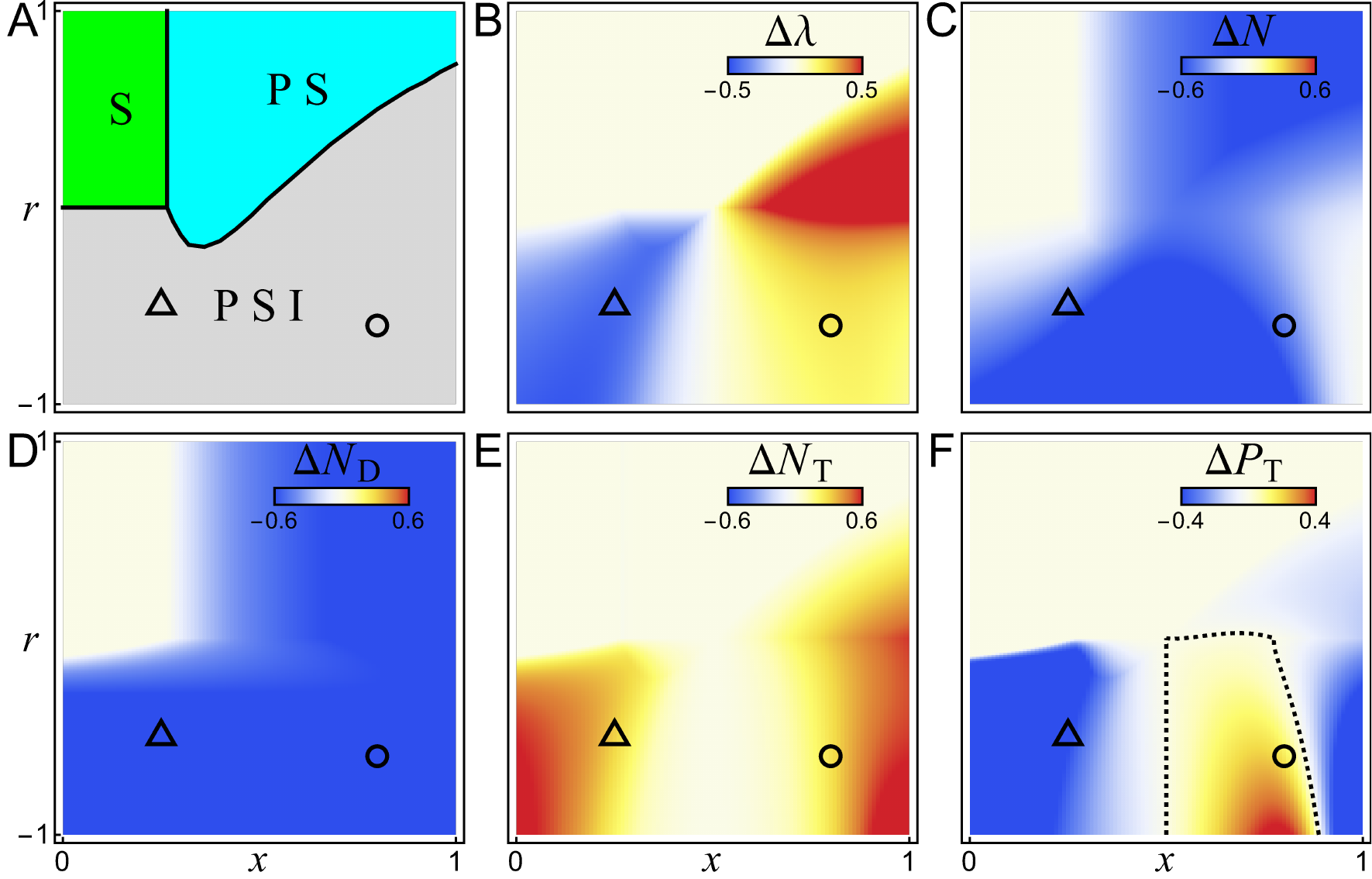}
\caption{The intimidated prey model. (A) Phase diagram showing the equilibrium of the system depending on the parameters $x$ (consumption preference of the predator) and $r$ (growth difference between prey types). The S-phase (green) is where only sensitive prey are present, the PS-phase (cyan) is where the predator coexists with only sensitive prey, and the PSI-phase (gray) is where the predator coexists with both sensitive and intimidated prey. (B,C) Heatmaps showing the change in the prey composition ($\Delta \lambda$) and density ($\Delta N$) after the predator is introduced. (D,E) Heatmaps showing the density and trait effects that make up the total population change, $\Delta N = \Delta N_D + \Delta N_T$. (F) Heatmap showing the back-reaction of the trait effect on the predator ($\Delta P_T$). Parameter values used in these plots are the same as for Figure~\ref{fig:time_all}AB.}
\label{fig:phase_int}
\end{figure}

\begin{figure}
\centering
\includegraphics[width=.95\textwidth]{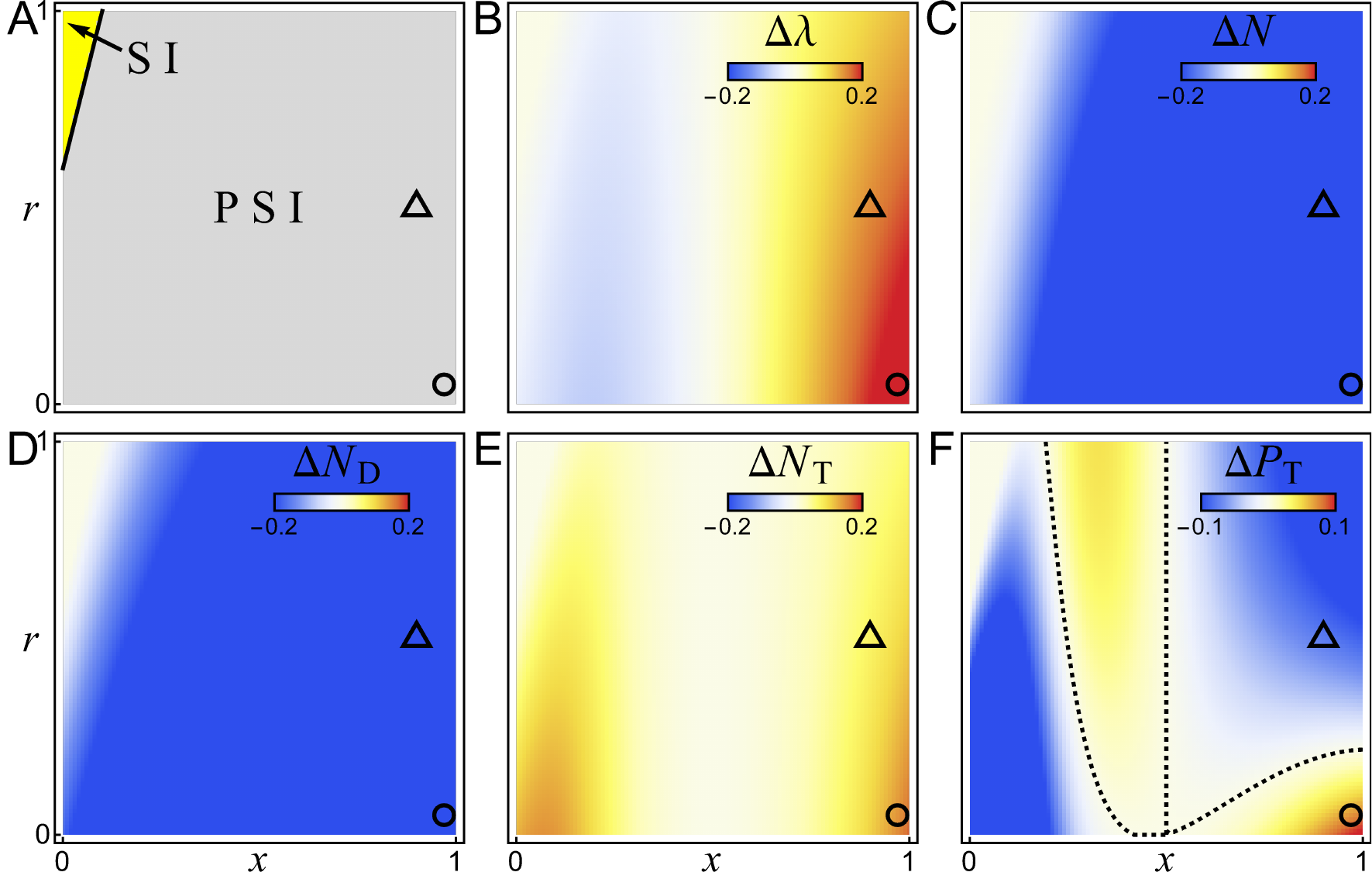}
\caption{The inherent heterogeneity model. (A) Phase diagram showing the equilibrium of the system, similar to Figure~\ref{fig:phase_int}A. The SI-phase (yellow) is where both superior and inferior prey types are present but the predator cannot persist, and the PSI-phase (gray) is where the predator coexists with both prey types. (B-F) Heatmaps showing the trait shift ($\Delta \lambda$), the change in prey population ($\Delta N$), the density and trait effects ($\Delta N_D$, $\Delta N_T$), and the back-reaction on the predator ($\Delta P_T$), similar to those in Figure~\ref{fig:phase_int}. Parameter values are the same as for Figure~\ref{fig:time_all}CD.}
\label{fig:phase_inh}
\end{figure}

\begin{figure}
\centering
\includegraphics[width=.95\textwidth]{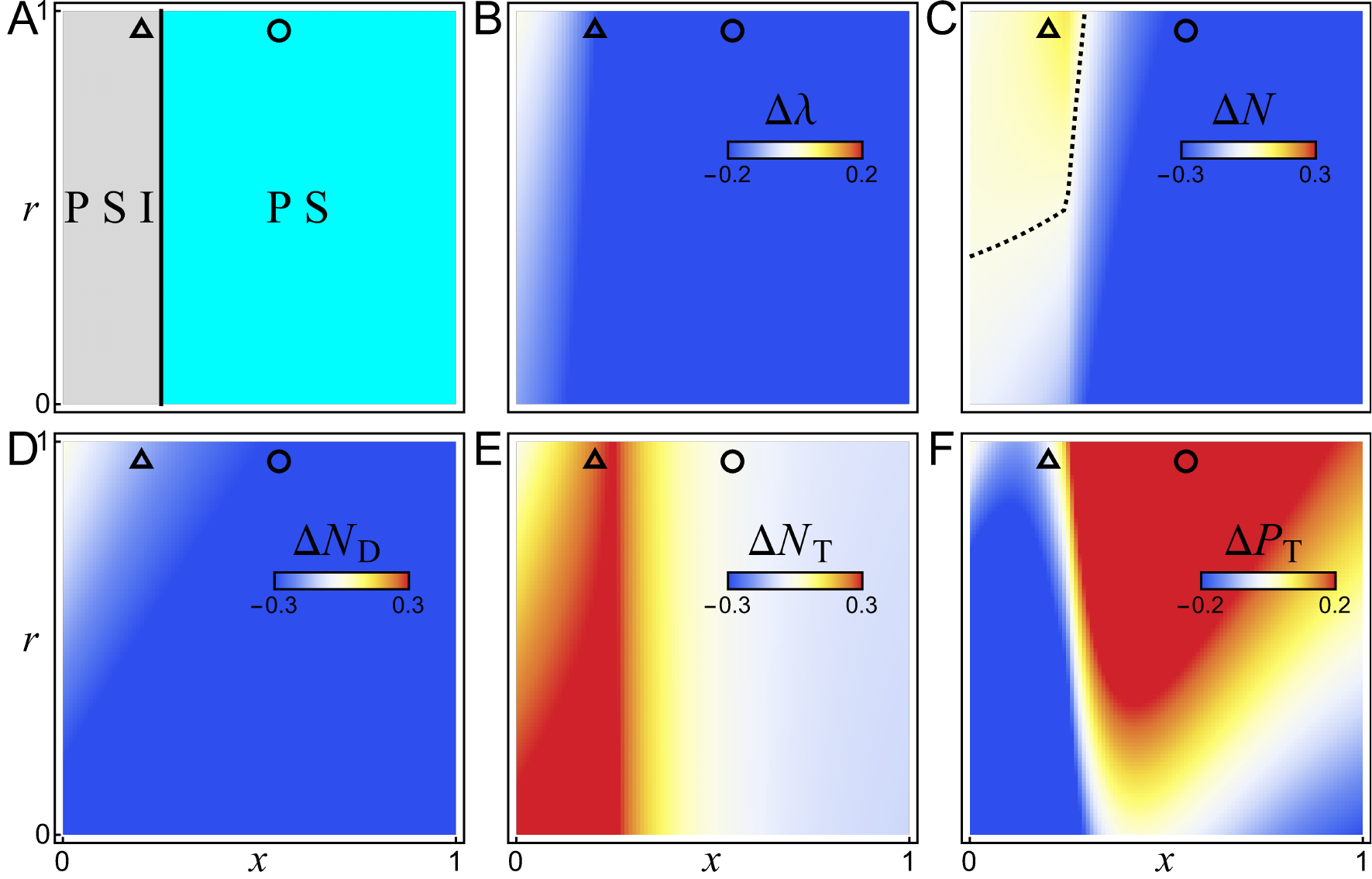}
\caption{The infectious disease model. (A) Phase diagram showing the equilibrium of the system, similar to Figure~\ref{fig:phase_int}A. The SI-phase (yellow) is where both susceptible and infected prey types are present but the predator cannot persist, the PS-phase (cyan) is where the predator coexists with only susceptible prey, and the PSI-phase (gray) is where the predator coexists with both susceptible and infected prey. (B-F) Heatmaps showing the trait shift ($\Delta \lambda$), the change in prey population ($\Delta N$), the density and trait effects ($\Delta N_D$, $\Delta N_T$), and the back-reaction on the predator ($\Delta P_T$), similar to those in Figure~\ref{fig:phase_int}. Parameter values are the same as for Figure~\ref{fig:time_all}EF.}
\label{fig:phase_inf}
\end{figure}

We present our results as a phase diagram that shows whether a population (or subpopulation) persists at the equilibrium (panel A in Figures~\ref{fig:phase_int}--\ref{fig:phase_inf}). ``P'' represents the predator, ``S'' represents type-1 prey (sensitive, superior, or susceptible), and ``I'' represents type-2 prey (intimidated, inferior, or infected). In the model of intimidated prey, three kinds of equilibria are possible: S, PS, and PSI (Figure~\ref{fig:phase_int}A). The S-phase is where only sensitive (unintimidated) prey are present, because the predator cannot persist. The PS-phase is where the predator persists but is not enough to intimidate the prey, so there is only the predator and sensitive prey. The PSI-phase is where the predator and both types of prey coexist. In the model of inherent heterogeneity, both types of prey are persistent because they are born in a given ratio. Consequently, there are two possible phases: SI and PSI (Figure~\ref{fig:phase_inh}A). The SI-phase is where the predator cannot persist because it strongly prefers the slowly growing prey type, and the PSI-phase is where the predator coexists with the prey. In the model of infectious disease, there are three phases: SI, PS, and PSI (Figure~\ref{fig:phase_inf}A). The SI-phase is where the predator cannot persist, similar to the other models, and the prey consists of both types because of the spreading disease. The PS-phase is where the predator persists but the infection does not spread, so all prey are healthy. Finally, the PSI-phase is where the predator, susceptible prey, and infected prey coexist.

We also plot heatmaps to show how the trait shift ($\Delta \lambda$), density effect ($\Delta N_D$) and trait effect ($\Delta N_T$) depend on the parameters (panels B,D,E in Figures~\ref{fig:phase_int}--\ref{fig:phase_inf}). There are parameter regimes for all three models where the effects of introducing the predator are as expected. These correspond to regions where the predator has a strong preference in prey type. In these regions the introduction of the predator shifts the trait value $\lambda$ away from its preference. This causes the trait effect $\Delta N_T$ to be positive, because the shift of the prey distribution reduces the overall consumption rate of the predator. On the other hand, the density effect $\Delta N_D$ is always negative, because the introduction of a predator without allowing a shift in the prey composition can only result in a positive consumption rate that reduces the prey population. In most cases the trait effect on the prey is not large enough to overcome the density effect from the predator, so the overall change in the prey population $\Delta N$ (panel C in Figures~\ref{fig:phase_int}--\ref{fig:phase_inf}) is negative. However, in some cases the trait effect is strong enough to change the result qualitatively, as we show below.

\subsection*{Emergent promotion of prey by predator}

In the region between the dashed lines in Figure~\ref{fig:phase_inf}C, we have $\Delta N > 0$, which means the prey population is larger when the predator is present rather than absent. This is an example of \textit{emergent promotion} described in \cite{jackson:2023}, whereby the introduction of a predator counterintuitively increases the prey population. Note that in the same region we have $\Delta N_D < 0$ (Figure~\ref{fig:phase_inf}D), which means the prey population would have decreased if the trait distribution was held fixed. Therefore, \textit{emergent promotion only arises because of the trait shift} in the prey. Here the trait shift is away from the predator's preference ($\Delta \lambda < 0$, Figure~\ref{fig:phase_inf}B), so the trait effect is positive ($\Delta N_T > 0$, Figure~\ref{fig:phase_inf}E). In this particular region, the positive trait effect is strong enough to overcome the negative density effect, resulting in a net increase in the prey population.

\begin{figure}
\centering
\includegraphics[width=.95\textwidth]{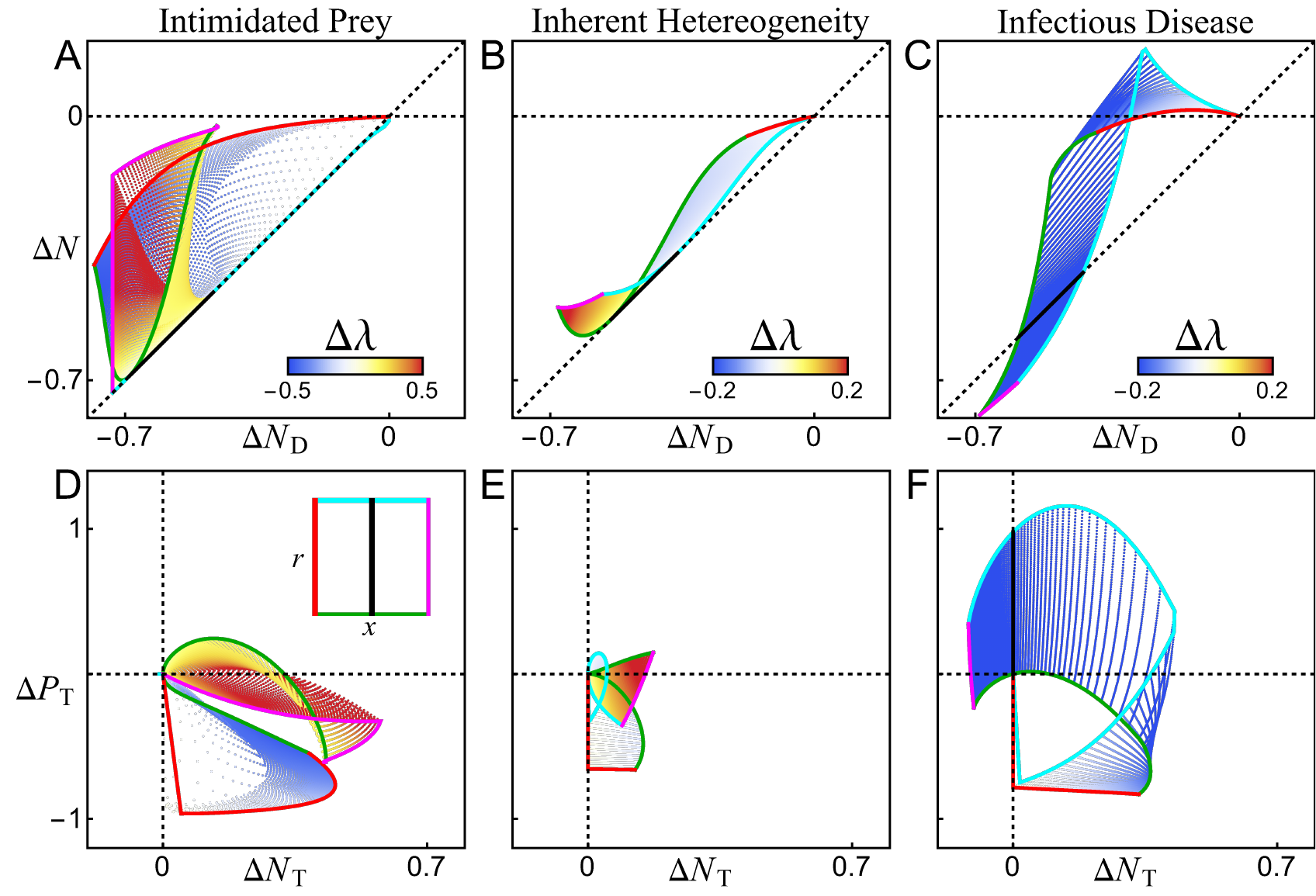}
\caption{(A--C) Relationship between the total effect of the predator on the prey ($\Delta N$) and the density effect ($\Delta N_D$). The diagonal dashed line is where the trait effect $\Delta N_T = 0$, and the horizontal dashed line is where $\Delta N = 0$. The density effect is always negative; above the diagonal is where the trait effect is positive, and above the horizontal line is where the trait effect is strong enough to overcome the negative density effect to result in emergent promotion of the prey by the predator. (D--F) Relationship between the trait effect on the prey ($\Delta N_T$) and the back-reaction on the predator ($\Delta P_T$). The back-reaction can be either negative or positive. The scattered points correspond to grid points in the parameter space from Figures~\ref{fig:phase_int}--\ref{fig:phase_inf}, colored according to $\Delta \lambda$; solid lines correspond to the boundaries of the parameter space from Figures~\ref{fig:phase_int}--\ref{fig:phase_inf}. $\Delta N$, $\Delta N_D$, and $\Delta N_T$ are normalized by the prey population before introducing the predator, and $\Delta P_T$ is normalized by the predator population without trait shift.}
\label{fig:cluster}
\end{figure}

The relationship between $\Delta N_D$ and $\Delta N_T$ can be visualized with a scatter plot shown in Figure~\ref{fig:cluster}. Each point corresponds to a particular pair of ($x$,$r$) values, and is colored by the value of $\Delta \lambda$ (from panel B in Figures~\ref{fig:phase_int}--\ref{fig:phase_inf}). In Figure~\ref{fig:cluster}ABC, the diagonal dashed line is where $\Delta N = \Delta N_D$ (i.e., $\Delta N_T = 0$), and the horizontal dashed line is where $\Delta N = 0$. In the lower triangle between the two dashed lines, the trait shift has a positive effect on the prey population ($\Delta N_T > 0$), but not enough to overcome the negative density effect ($\Delta N_D < 0$), and so the overall effect of the predator is to reduce the prey population ($\Delta N < 0$). Above the horizontal dashed line, the effect of the trait shift is large enough to overcome the density effect (so that $\Delta N > 0$), leading to the promotion of the prey by the predator. Note that for all three models that we consider, $\Delta N_D$ is always negative. That is, in the absence of a trait shift, introducing a predator always decreases the population of the prey. For both the intimidated prey and inherent heterogeneity models (Figure~\ref{fig:cluster}AB), all points are inside the lower triangle between the dashed lines, meaning that the trait effect is always positive but not enough to cause promotion. However, for the infectious disease model (Figure~\ref{fig:cluster}C), there are points above the horizontal dashed line, showing that the effect of the trait shift can be strong enough to promote the prey population overall.

\subsection*{Back-reaction on predator}

In most regions of the phase diagrams, particularly where the predator has a strong preference, the trait shift is towards a prey composition that is unfavorable for the predator. This can be understood as a result of picking up all the ``lower-hanging fruits'', only to be left with hard-to-get targets. In analogy, some prey are consumed more readily due to their particular trait values, and the remaining ones will be composed of more unfavorable prey. This forms a negative back-reaction on the predator ($\Delta P_T < 0$), as the predator is responsible for shifting the prey composition unfavorably for itself. Some regions of the parameter space that show this effect are the regions containing the triangular points in Figures~\ref{fig:phase_int}--\ref{fig:phase_inf}.

However, there are situations where the back-reaction is positive for the predator ($\Delta P_T > 0$). One region of particular interest is outlined in Figure~\ref{fig:phase_inh}F, enclosing the circular point. Here the trait shift is strongly away from the predator's preference, but $\Delta P_T$ is still positive. This is because there are two competing factors at play. The first is that the trait shift in the prey causes the predator to consume at a lower average rate. The second is that the trait shift also increases the equilibrium prey population ($\Delta N_T > 0$), which helps increase the predator population. If the latter factor is more dominant, then it leads to the surprising result that the trait shift goes against the predator's preference but is favorable for the predator after all.

It is also interesting that in the above case the trait shift benefits both the predator and prey at the same time. Figure~\ref{fig:cluster}DEF visualizes the relationship between $\Delta P_T$ and $\Delta N_T$. The upper right quadrant is where both $\Delta P_T > 0$ and $\Delta N_T > 0$. In the models of intimidated prey and inherent heterogeneity (Figure~\ref{fig:cluster}DE), even though the back-reaction on the predator can be either positive or negative, the trait effect on the prey is always positive ($\Delta N_T > 0$). In the infectious disease model (Figure~\ref{fig:cluster}F), there are cases where the trait effect on the prey is negative ($\Delta N_T < 0$), because in this scenario the trait shift not only results from the predator through consumption or intimidation, but also follows its own dynamic rules (the infection can be considered a parasite species that has its own population dynamics).

\section*{Discussion}

Our analysis of the above models in a unified framework represents a general way of treating interaction between the predator and prey when there is intraspecific trait variation. We have found qualitatively different, and sometimes surprising, behaviors of these systems when the dynamic shift of the trait distribution is taken into account. Those results can be understood by comparing the relative strengths of density and trait effects from the predator to the prey, as well as the back-reaction to the predator.

\subsection*{Direct vs indirect trait-mediated effects}

The interaction strength between a pair of species can be affected by the presence of other species in their environment. When a third species affects the trait of one of the interacting species and thus modifies the interaction strength between those two species, it is known as a ``trait-mediated indirect effect'' \cite{holt:2012}. Past studies of trait effects have mostly focused on such indirect effects, such as in trophic cascades \cite{huang:1991, wissinger:1993, grabowski:2005, wodjak:2005}. In these situations, the predator may affect the behavior of the prey, thereby alter the prey's consumption of lower level resources. The trait effect that we focus on in our model can be called a ``direct'' effect, because it is a modification of the interaction strength by one of the two interacting species. This may be the simplest form of trait-mediated effects involving the fewest number of species.

\subsection*{Healthy herd effect}

When the introduction of a predator reduces the prevalence of a disease in a prey population, it is known as the ``healthy herd effect'' \cite{packer:2003,lopez:2021,lopez:2023,gallagher:2019}. This effect can be seen in the $\Delta \lambda$ plot for the infectious disease model (Fig.~\ref{fig:phase_inf}B). Since here $\lambda$ represents the proportion of infected prey, the healthy herd effect is present wherever $\Delta \lambda < 0$. However, there can be different mechanisms behind this effect \cite{duffy:2011}. It can arise from either a reduction in prey population density that reduces the spread of the disease \cite{lopez:2023,gallagher:2019}, or a strong preference for infected prey by the predator that reduces the subpopulation of infected prey \cite{duffy:2011}. Our results show clear regions in the parameter space that represent each of these mechanisms. In the region of $x > 0.5$, the predator prefers healthy prey, so the effect must be from the reduction in prey density that leads to a slower infection rate. On the other hand, in the region enclosed in the dashed curve (Fig.~\ref{fig:phase_inf}C), the prey density actually increases upon introducing the predator, and this is also a region where the predator strongly prefers to consume infected prey, so the effect must be from preferential consumption. In between these two regions, there is a continuum where the healthy herd effect is caused by a combination of the two mechanisms. Moreover, we can distinguish these two mechanisms by whether the density effect or the trait effect dominates. Where the density effect is strong compared to the trait effect (Fig.~\ref{fig:phase_inf}D), it is the reduction in density that causes a healthier prey population; where the trait effect is dominant (Fig.\ref{fig:phase_inf}E), it is the preferential consumption by the predator that causes the healthy herd effect.

\subsection*{Alternative definitions of trait effect}

Our way of splitting the overall effect of a predator on the prey into density and trait effects is ideal for situations where the trait distribution $\lambda$ can be measured and controlled. To estimate $\Delta N_T$, for example, requires measuring the predator and prey populations while holding $\lambda$ constant. This is not always practical. There are alternative definitions of density and trait effects (as reviewed in \cite{okuyama:2007,holt:2012}), which focus on the indirect effect of the predator on lower level resources. Specifically, one defines a ``true predator'' that has all its natural interactions with the prey, and a ``threat predator'' that lacks the ability to consume the prey but can still influence the prey behavior by their presence. These definitions are useful for experiments such as \cite{peacor:2001,griffin:2006}, where the predator can be disarmed or caged to remove their consumption ability. In those experiments, the trait effect is defined as $N^{\dagger} - N_0$, and the density effect is $N^* - N^\dagger$, where $N^\dagger$ is the equilibrium density of the prey in the presence of the threat predator, and $N^*$ and $N_0$ are the same as in our definition (the equilibrium prey density with and without the true predator). 

Compared to our definitions of density and trait effects, the difference is in the choice of the midpoint and the order of subtraction. Whereas we use $N^*(\lambda_0)$ as the midpoint to parse out the density effect first, the alternative definitions above use $N^\dagger$ to parse out the trait effect first. The latter removes the need to control the trait composition of the prey directly. However, it also introduces ambiguity into interpreting the density effect defined this way. This is because the threat predator and true predator will not necessarily shift the trait of the prey in the same way. For example, the true predator may further shift the prey traits by preferential consumption, as in our model. Thus, the density effect so defined will include some unknown amount of trait effect. In contrast, with the definitions that we propose, no such ambiguity exists, at the expense of having to control the trait distribution directly. 

Such control has been implemented in \cite{peacor:2001,griffin:2006} by ``culling'' the prey, i.e., a treatment where no predator is present but some prey are removed manually to simulate the action of predator without affecting the behavior of the prey. This is similar to our definition of $N^*(\lambda_0)$ in that it attempts to isolate the density effect of the predator without inducing a trait shift. This protocol is useful when investigating the \textit{indirect} effect on lower level resources. However, if the goal is to measure the \textit{direct} effect on the prey itself, the prey population should be measured rather than manipulated; only the composition of the prey should be held constant. This could potentially be accomplished by replacing individuals in the population with those of particular trait values from a separate pool to maintain a constant trait composition in the population. Such a replacement protocol has been used in eco-evolutionary studies to constrain the evolution of populations in multi-generational studies \cite{pimentel:1963,szHucs:2017}.

We have also defined a back-reaction of the trait effect on the predator population. Previous experiments on trait effects are often short-term, during which the predator population remains constant. Consequently, no back-reaction on the predator can be measured. To test the trait-mediated back-reaction in our model would require longer-term experiments where the predator population is allowed to equilibrate, alongside the prey population. Our theoretical results shown here provide an expectation for such longer-term empirical studies in future.

\section*{Conclusion}

We have presented simple predator-prey models to demonstrate the effect that a dynamic trait distribution within a population can have on the interaction between species. Such effects can have profound implications for larger, more diverse ecosystems. For instance, in the context of trophic cascade, it is often thought that a predator would have a negative impact on its immediate prey, but a positive effect on species at the next trophic level \cite{pace:1999}. However, emergent promotion that is made possible by dynamic trait shift, as illustrated in our model, can potentially reverse the expected trend. By analyzing the density and trait effects, we expect to be able to understand such unintuitive outcomes of community dynamics and avoid unwarranted assumptions about species interactions.

Our work shows the importance of considering the dynamical shift of trait distributions on short timescales due to ecological processes, complementary to the slow shift due to evolution that has been studied in eco-evolutionary models \cite{ferriere:2015, govaert:2019, pastore:2021, schreiber:2011}. The models studied here are simplified examples that include only two trait values within a single species. This approach can be generalized by considering more compartments or a continuous distribution of trait values within a population. We have considered the shift of a single trait, but in reality even a single species has many traits that can shift and alter its interactions in a large ecosystem. How to analyze realistic systems with many species, each with a high-dimensional space of shifting traits, is left to future study.

\section*{Acknowledgments}

We thank Robert Holt and Mathew Leibold for helpful discussions.

\bibliographystyle{unsrt}

\pagebreak

\appendix

\section*{Appendix A: Additional Methods }\

\renewcommand{\theequation}{A\arabic{equation}}
\renewcommand{\thetable}{A\arabic{table}}
\setcounter{equation}{0} 
\setcounter{table}{0}

\subsection*{Figure production}

The time series shown in Figure~\ref{fig:time_all} were produced by integrating the corresponding system of equations using the function \texttt{NDSolve} in Mathematica 13.2. The initial values shown at $t=0$ in the plots are the equilibrium values for $N$ and $\lambda$ in the absence of the predator, found by integrating the system with $P=0$ for 200 dimensionless time units. The hypothetical cases of a fixed trait distribution were generated by integrating the system with $\lambda$ held constant at its initial value. In all cases we studied the system has a single stable equilibrium. For consistency, we verified that if the constraint on $\lambda$ in the hypothetical cases is relaxed after reaching equilibrium, the dynamics will take the system to the equilibrium of the full system.

The phase diagrams that are panel A in Figures~\ref{fig:phase_int}--\ref{fig:phase_inf} were produced by integrating the corresponding system of equations using \texttt{NDSolve} for 1000 dimensionless time units and then checking whether the abundance of each (sub)population is above a threshold of 0.0001. Panels B--F in Figures~\ref{fig:phase_int}--\ref{fig:phase_inf} were produced by subtracting the equilibrium populations of the corresponding systems as described in the Methods section. In cases where the predator is excluded, the equilibrium is calculated analytically by solving for where the time derivatives are equal to zero. In cases where the predator is present and $\lambda$ is dynamic, the equilibrium is found by integrating the system using \texttt{NDSolve} for 1000 dimensionless time units. The stability of the equilibrium is verified by calculating the Jacobian of the system at the equilibrium and checking that its eigenvalues all have negative real parts. Each of these calculations is done on a grid in the parameter space with an increment of $0.01$, and the results are plotted using the Mathematica function \texttt{ListDensityPlot}.

The scatter plots in Figure~\ref{fig:cluster} are produced with the same data shown in  Figures~\ref{fig:phase_int}-\ref{fig:phase_inf}, normalized, then plotted using the Mathematica function \texttt{ListPlot}.

\end{document}